\documentclass{ws-ijmpa}

\begin{document}

\markboth{Masayasu Harada}{%
Vector Manifestation and the Hidden Local Symmetry}

\catchline{}{}{}{}{}

\title{%
\hfill{\footnotesize
\vbox{\hbox{\rm DPNU-05-17}\hbox{\rm September, 2005}  }}\\
\vspace{0.1cm}
Vector Manifestation and the Hidden Local Symmetry\footnote{%
Talk given at
``International Conference on QCD and Hadronic Physics''
(June 16-20, 2005, Beijing, China).
}
\vspace{-0.3cm}
}

\author{Masayasu Harada}
\address{Department of Physics, Nagoya University,
Nagoya 464-8602, Japan\footnote{
Electronic address : harada@hken.phys.nagoya-u.ac.jp}
}


\maketitle

\begin{abstract}
In this write-up, I summarize the key ingredients
of the vector manifestaion formulated in the
hidden local symmetry theory, in which the $\rho$ meson
becomes massless degenerated with the pion at the
chiral phase transition point.
\end{abstract}



\section{Introduction}

The vector manifestation (VM) was proposed~\cite{HY:VM,HY:PRep}
as a novel manifestation of Wigner 
realization of
chiral symmetry where the vector meson $\rho$ becomes massless at the
chiral phase transition point. 
Accordingly, the (longitudinal) $\rho$ becomes the chiral partner of
the Nambu-Goldstone boson $\pi$.
The VM provides a strong support for
Brown-Rho scaling~\cite{BR} which predicted that the mass of
light-quark hadrons should drop in proportion to the quark
condensate $\langle\bar{q}q\rangle$.
Actually, the dilepton spectra observed in 
the KEK-PS E325~\cite{KEK-PS} and
the CBELSA/TAPS~\cite{trnka}
as well as in the CERN/SPS~\cite{ceres}
can be explained by the 
dropping mass of $\rho$/$\omega$ based on the Brown-Rho
scaling~\cite{LKB-RW}.

In this write-up, I first briefly review the difference between the 
VM and
the conventional manifestation of chiral symmetry restoration
based on the linear sigma model in section~\ref{sec:VM}.
Then, I show how to formulate the VM in hot matter in 
section~\ref{sec:VMT}.
Finally, in section~\ref{sec:sum},
I give a brief summary.

\section{Vector Manifestation of Chiral Symmetry}
\label{sec:VM}

Let me discuss the difference between the 
VM and
the conventional manifestation of chiral symmetry restoration
based on the linear sigma model,
the GL manifestation,
in terms of the chiral representation of the 
low-lying mesons
by extending the analyses done in
Ref.~\refcite{Gilman-Harari-Weinberg:69}
for two flavor QCD.

The VM is characterized by
\begin{equation}
\mbox{(VM)} \qquad
f_\pi^2 \rightarrow 0 \ , \quad
m_\rho^2 \rightarrow m_\pi^2 = 0 \ , \quad
f_\rho^2 / f_\pi^2 \rightarrow 1 \ ,
\label{VM def}
\end{equation}
where $f_\rho$ is the decay constant of 
(longitudinal) $\rho$ at $\rho$ on-shell.
This is completely different from 
the GL manifestation
where the scalar meson $S$ becomes massless
degenerate with $\pi$ as the chiral partner

I first consider 
the representations of 
the following zero helicity ($\lambda=0$) states
under
$\mbox{SU(3)}_{\rm L}\times\mbox{SU(3)}_{\rm R}$;
the $\pi$, the (longitudinal) $\rho$, 
the $S$ and
the (longitudinal) axial-vector
meson denoted by $A_1$.
The $\pi$ and the $A_1$ 
are admixture of $(8\,,\,1) \oplus(1\,,\,8)$ and 
$(3\,,\,3^*)\oplus(3^*\,,\,3)$
since the symmetry is spontaneously
broken~\cite{Gilman-Harari-Weinberg:69}:
\begin{eqnarray}
\vert \pi\rangle &=&
\vert (3\,,\,3^*)\oplus (3^*\,,\,3) \rangle \sin\psi
+
\vert(8\,,\,1)\oplus (1\,,\,8)\rangle  \cos\psi
\ ,
\nonumber
\\
\vert A_1(\lambda=0)\rangle &=&
\vert (3\,,\,3^*)\oplus (3^*\,,\,3) \rangle \cos\psi 
- \vert(8\,,\,1)\oplus (1\,,\,8)\rangle  \sin\psi
\ ,
\label{mix pi a}
\end{eqnarray}
where the experimental value of the mixing angle $\psi$ is 
given by approximately 
$\psi=\pi/4$~\cite{Gilman-Harari-Weinberg:69}.  
The $\rho$ and the $S$, 
on the other hand, are expressed as
\begin{eqnarray}
\vert \rho(\lambda=0)\rangle =
\vert(8\,,\,1)\oplus (1\,,\,8)\rangle  
\ ,
\quad
\vert S\rangle 
=
\vert (3\,,\,3^*)\oplus (3^*\,,\,3) \rangle 
\ .
\label{rhos}
\end{eqnarray}

When the chiral symmetry is restored at the
phase transition point, 
it is natural to expect that
the chiral representations coincide with the mass eigenstates:
The representation mixing is dissolved.
{}From Eq.~(\ref{mix pi a}) one can easily see
that
there are two ways to express the representations in the
Wigner phase of the chiral symmetry:
The conventional GL manifestation
corresponds to 
the limit $\psi \rightarrow \pi/2$:
\begin{eqnarray}
\mbox{(GL)}
\qquad
\left\{
\begin{array}{rcl}
\vert \pi\rangle\,, \vert S\rangle
 &\rightarrow& 
\vert  (3\,,\,3^\ast)\oplus(3^\ast\,,\,3)\rangle\ ,
\\
\vert \rho (\lambda=0) \rangle \,,
\vert A_1(\lambda=0)\rangle  &\rightarrow&
\vert(8\,,\,1) \oplus (1\,,\,8)\rangle\ .
\end{array}\right.
\end{eqnarray}
On the other hand, the VM corresponds 
to the limit $\psi\rightarrow 0$ in which the $A_1$ 
goes to a pure 
$(3\,,\,3^*)\oplus (3^*\,,\,3)$, now degenerate with
the scalar meson $S$ in the same representation, 
but not with $\rho$ in 
$(8\,,\,1)\oplus (1\,,\,8)$:
\begin{eqnarray}
\mbox{(VM)}
\qquad
\left\{
\begin{array}{rcl}
\vert \pi\rangle\,, \vert \rho (\lambda=0) \rangle
 &\rightarrow& 
\vert(8\,,\,1) \oplus (1\,,\,8)\rangle\ ,
\\
\vert A_1(\lambda=0)\rangle\,, \vert s\rangle  &\rightarrow&
\vert  (3\,,\,3^\ast)\oplus(3^\ast\,,\,3)\rangle\ .
\end{array}\right.
\end{eqnarray}
Namely, the
degenerate massless $\pi$ and (longitudinal) $\rho$ at the 
phase transition point are
the chiral partners in the
representation of $(8\,,\,1)\oplus (1\,,\,8)$.

\section{Formulation of the Vector Manifestation in Hot Matter}
\label{sec:VMT}

It should be noticed that
the critical temperature of the
chiral symmetry restoration is approached 
from the broken phase up to $T_c - \epsilon$,
and that
the following basic assumptions are adopted in the 
analysis:
(1) The relevant degrees of freedom until near $T_c - \epsilon$
are only $\pi$ and $\rho$;
(2) Other mesons such as $A_1$ and scalar mesons are still heavy 
at $T_c - \epsilon$;
(3) Partial chiral symmetry restoration already occurs at 
$T_c - \epsilon$.
Based on these assumptions, 
it was shown~\cite{HS:VMT,HS:VD,Sasaki:thesis}
that the VM 
necessarily occurs at the chiral symmetry restoration point.

One of the most important
key ingredients to formulate the VM in hot matter 
is
the {\it intrinsic temperature
dependences}~\cite{HS:VMT}
of the parameters determined through the Wilsonian
matching~\cite{HY:WM}.
The intrinsic temperature dependence introduced through
the Wilsonian matching is
nothing but the signature that hadrons have internal structures
constructed from the quarks and gluons.
This is similar to the situation where coupling constants
among hadrons are replaced with appropriate
momentum-dependent form factors
in high energy region.
The intrinsic effects play more important
roles in higher temperature region, 
especially near the critical temperature.

When $T_c$ is approached from below,
the axial-vector and vector current correlators
derived in the OPE approach each other
for any value of $Q^2$.
Thus we require that
these current correlators in the HLS become close to each other
near $T_c$
for any value of $Q^2\ \mbox{around}\ {\Lambda}^2$.
This requirement implies that
the bare $g$ and $a$ satisfy
\begin{eqnarray}
&&
g_{\rm bare}(T) \mathop{\longrightarrow}_{T \rightarrow T_c} 0 \ ,
\qquad
a_{\rm bare}(T) \mathop{\longrightarrow}_{T \rightarrow T_c} 1 \ .
\label{g a :VMT}
\end{eqnarray}
These conditions (``VM conditions in hot matter'')
for the bare parameters
are converted into the
conditions for the on-shell parameters through the Wilsonian 
renormalization group equations (RGEs).
Since $g=0$ and $a=1$ are separately the fixed points of the RGEs for
$g$ and $a$~\cite{HY:letter},
$(g,a)=(0,1)$ is satisfied at any energy scale.
As a result, the quantum correction to the $\rho$ mass
as well as the hadronic thermal correction disappears at $T_c$
since they are proportional to the gauge coupling $g$.
The bare $\rho$ mass, which is also proportional to $g_{\rm bare}$,
vanishes at $T_c$.
These imply that the pole mass of the $\rho$ meson also vanishes
at $T_c$:
\begin{eqnarray}
&&
m_\rho^2(T)
\rightarrow 0 \ \ \mbox{for} \ T \rightarrow T_c \ .
\end{eqnarray}

\vspace{-0.3cm}

\section{Summary}
\label{sec:sum}

In this write-up,
I first showed the difference between the 
VM and
the conventional manifestation of chiral symmetry restoration
based on the linear sigma model
in terms of the chiral representation of the mesons
in section~\ref{sec:VM}.
Then, in section~\ref{sec:VMT}, I
reviewed how to formulate the VM in hot matter.
I would like to note
that the VM in dense matter can be formulated
in a similar way~\cite{HKR}, 
where the {\it intrinsic density dependence}
plays an important role.

There are several predictions of the VM in hot matter
made so far:
In Ref.~\refcite{HKRS},
the vector and axial-vector susceptibilities were studied.
It was shown that the equality between two susceptibilities
are satisfied and that the VM predicts
$\chi_A = \chi_V = \frac{2}{3} \, T_c^2$ for $N_f = 2$,
which is in good agreement with the result obtained in the lattice 
simulation~\cite{Allton}.
In Ref.~\refcite{HS:VD},
a prediction associated with the validity of 
vector dominance (VD) in hot matter was made:
As a consequence of including the intrinsic effect,
the VD is largely violated at the critical temperature.
The violation of the VD near the critical temperature
plays an important role to explain the NA60 dimuon data
based on the Brown-Rho scaling~\cite{BR:05}.
In Ref.~\refcite{VM:Vpi},
the pion velocity $v_\pi$ was studied
including the effect of Lorentz symmetry 
breaking.
The prediction for $v_\pi$ is $v_\pi (T_c)= 0.83 - 0.99$,
which is 
consistent with $v_\pi(T) = 0.65$ of 
Cramer {\it et al.}~\cite{Cramer} extracted from
the recent STAR data~\cite{star}.
Furthermore, in Ref.~\refcite{HRS:D},
starting with an HLS Lagrangian at the 
VM fixed point that incorporates the heavy-quark
symmetry and matching the bare theory to QCD, we calculated the
splitting of chiral doublers of $D$ mesons proposed in
Refs.~\cite{NRZ-BH}.
The predicted splitting 
comes out to be $0.31\pm0.12$\,GeV,
which is in good agreement with the 
experiment~\cite{BABAR-CLEO-Belle}.


\section*{Acknowledgements}

I would like to thank 
Doctor Youngman Kim, Professor Mannque Rho,
Doctor Chihiro Sasaki and
Professor Koichi Yamawaki for collaboration in
several works done for the vector manifestaion
on which this talk is based.
This work is supported in part by
the JSPS Grant-in-Aid for Scientific Research (c) (2) 16540241,
the Daiko Foundataion \#9099,
and 
the 21st Century COE
Program of Nagoya University provided by Japan Society for the
Promotion of Science (15COEG01).


\end{document}